# Photo-induced charge state dynamics of the neutral and negatively charged silicon vacancy centers in room-temperature diamond

*G. Garcia-Arellano, G. I. López-Morales, N. B. Manson, J. Flick, A. A. Wood, and C. A. Meriles*

The silicon vacancy (SiV) center in diamond is drawing much attention due to its optical and spin properties, attractive for quantum information processing and sensing. Comparatively little is known, however, about the dynamics governing SiV charge state interconversion mainly due to challenges associated with generating, stabilizing, and characterizing all possible charge states, particularly at room temperature. Here, we use multi-color confocal microscopy and density functional theory to examine photo-induced SiV recombination — from neutral, to single-, to double-negatively charged — over a broad spectral window in chemical-vapor-deposition diamond under ambient conditions. For the SiV$^0$ to SiV$^-$ transition, we find a linear growth of the photo-recombination rate with laser power at all observed wavelengths, a hallmark of single photon dynamics. Laser excitation of SiV$^-$, on the other hand, yields only fractional recombination into SiV$^{2-}$, a finding we interpret in terms of a photo-activated electron tunneling process from proximal nitrogen atoms.

## 1. Introduction

Optically active spin qubits in semiconductor materials have emerged as attractive candidates for quantum information processing and sensing due to the high level of control achievable over single and coupled spins in a variety of solid-state hosts[1-3]. Among the most promising and widely studied systems are color centers in diamond, the nitrogen-vacancy (NV) and silicon-vacancy (SiV) centers arguably being the best-known examples. The negatively charged NV (here denoted as NV$^-$) offers long spin coherence times at room temperature and is thus optimally suited for spin-based sensing[4,5] but its broad fluorescence spectrum renders it impractical in quantum protocols requiring the generation of identical photons. By contrast, the negatively charged SiV (or SiV$^-$) possesses narrow optical emission and high quantum efficiencies[6-9] — desirable for integration into photonic systems[10-13] — but its spin properties are poor. Bridging the gap between the two is the neutral silicon vacancy center (SiV$^0$), simultaneously combining narrow optical emission (with zero phonon line at 946 nm) and spin coherence times approaching one second at cryogenic temperatures[14-17]. The attractive optical and spin features of SiV$^0$ are tempered by challenges associated with generating and stabilizing the neutral charge state, typically requiring careful materials engineering of the host diamond. Formation of SiV$^0$ can be favored by shifting the Fermi level by chemical means, either by boron doping during crystal growth, or through hydrogen termination of the diamond surface[15,17,18]; an alternative possibility is to use above-bandgap UV excitation[19]. Exactly how the optical or spin properties of the defect are modified by these processes is generally not clear. Recently it was shown that SiV$^0$ can be generated in pristine chemical-vapor-deposition (CVD) diamond by two-step capture of holes diffusing from remote NV centers subjected to charge cycling under green excitation[20,21]. This approach is convenient in that it allows one to examine the charge dynamics of SiV$^0$ in the absence of the additional complexity created by proximity to dopants or the host crystal surface.

Here, we first investigate the population dynamics of SiV$^0$ centers with wavelengths in the range 720 to 874 nm. While SiV$^0$ is optically dark at room temperature, we infer its charge state dynamics by measuring the SiV$^-$ fluorescence levels upon SiV$^0$ recombination. For all studied wavelengths, we find a linear growth of the

G. Garcia Arellano, G. I. López Morales, J. Flick, C.A. Meriles
Department. of Physics,
CUNY-City College of New York,
New York, NY 10031, USA.
E-mail: cmeriles@ccny.cuny.edu

J. Flick, C.A. Meriles
CUNY-Graduate Center,
New York, NY 10016, USA.

J. Flick
Center for Computational Quantum Physics,
Flatiron Institute, New York, NY 10010, USA.

A. A. Wood
School of Physics,
The University of Melbourne,
Parkville VIC 3010 Australia.

N. B. Manson
Department of Quantum Science and Technology,
Research School of Physics,
Australian National University,
Canberra, A.C.T. 2601, Australia.

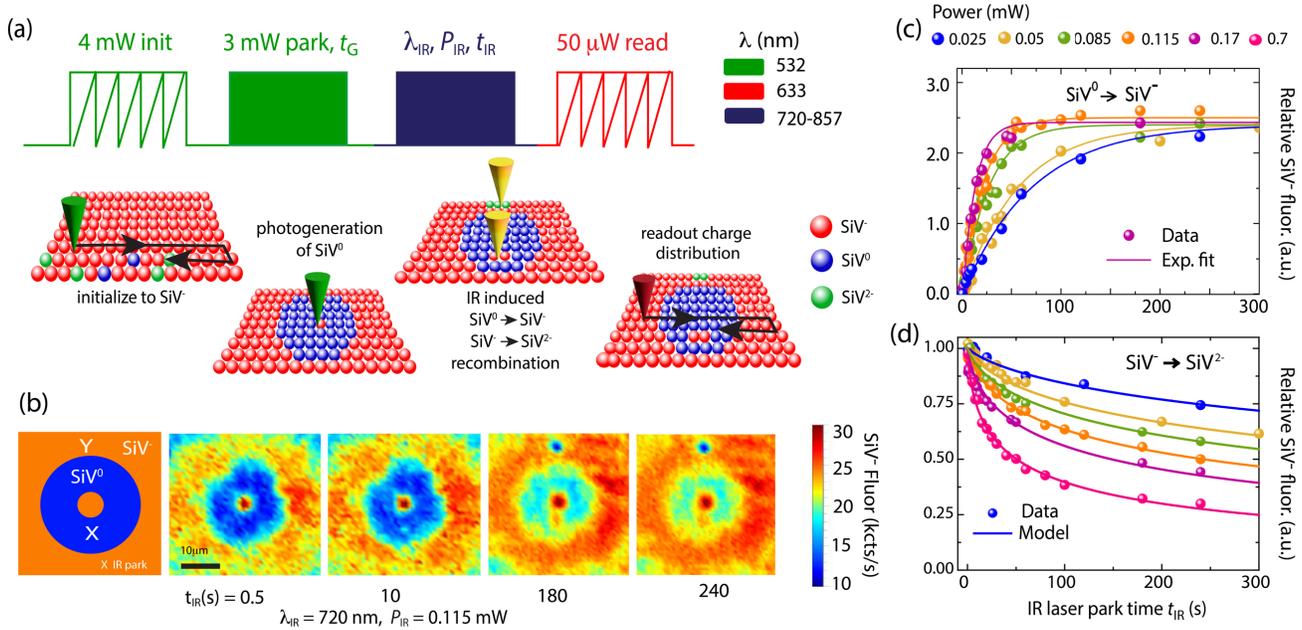

**Figure 1: Initialization and detection of SiV$^0$ and SiV$^-$ recombination under IR illumination.** (a) Experimental protocol. Following a 532 nm laser scan to produce an SiV$^-$-rich background (5 mW, 1 ms dwell time per pixel, green zigzag in the diagram), we generate an SiV$^0$-rich halo via a 3 s, 532 nm park (solid green block). Setting a tunable Ti:Sa laser to a variable wavelength $\lambda_{IR}$ and power $P_{IR}$, we successively park the beam at two locations, X and Y, respectively in the SiV$^0$- and SiV$^-$-rich areas of the crystal for a variable time $t_{IR}$ (solid blue block). Lastly, we image the resulting charge distribution via a weak red laser scan (632 nm at 50 μW, red zigzag). (b) Example confocal images upon application of the protocol in (a) for variable time $t_{IR}$. In this instance, $\lambda_{IR} = 720$ nm and $P_{IR} = 0.115$ mW; each image is an averaged composite of 4 scans per park time to account for imperfect initialization and power drifts of the Ti:Sa laser beam (<5%). (c) Integrated SiV$^-$ fluorescence (solid circles) at point X as a function of park time $t_{IR}$ for variable laser powers under 720 nm excitation; solid lines are exponential fits. (d) Integrated SiV$^-$ fluorescence (solid circles) versus $t_{IR}$ at point Y; solid lines represent fits to a model of nitrogen-assisted electron tunneling (see below). In (c) and (b), a.u.: arbitrary units.

recombination rate with laser power, indicative of a single photon process. Informed by density functional theory (DFT), we interpret the SiV$^0 \leftrightarrow$ SiV$^-$ recombination as the photo-injection of an electron from the diamond valence band into the SiV$^0$ excited state. For the same infra-red (IR) excitation window, we also study the dynamics of the negatively charged silicon-vacancy center as it transitions to the dark charge state corresponding to SiV$^{2-}$. Unlike SiV$^0$, we find this transformation manifests as a non-exponential time evolution of the observed fluorescence, difficult to account for in a framework of an isolated color center. Instead, numerical and *ab initio* modeling hint at a more complex process where photoexcitation of SiV$^-$ stimulates electron tunneling from a proximal substitutional nitrogen (N) impurity.

## 2. Experimental protocol

The sample we examine here is a [100] CVD-grown diamond with a nitrogen concentration of 3 ppm[22] and estimated SiV and NV concentrations around 0.3 and 0.03 ppm respectively. Along with nitrogen, silicon is a common contaminant in lower-grade CVD diamond, typically incorporated during crystal growth due to etching of the reactor quartz windows. Throughout our experiments, we use a custom-made multi-color scanning confocal microscope with 532 and 632 nm laser paths for excitation and detection of SiV centers. A tunable Ti:Sa laser running in continuous wave (cw) mode serves as third laser source for illumination at variable IR wavelengths in the range 700-1000 nm. We use a band pass filter to limit photon collection to a narrow window around the SiV$^-$ zero phonon line (ZPL) at 737 nm (see Supplementary Material, Section 1 for additional experimental details).

Characterizing the charge dynamics of SiV centers under optical excitation is challenging because, out of the three SiV charge states found in most CVD diamond samples — namely the neutral, single-, and double-negatively charged states — only SiV$^-$ exhibits significant room-temperature photoluminescence (in the form of a sharp peak at 737 nm and a weak phonon sideband[23]). Unlike the NV center, where absence of NV$^-$ fluorescence indicates NV$^0$ presence with near unitary fidelity[24], careful steps must be undertaken to unambiguously differentiate the optically inactive SiV$^{2-}$ from SiV$^0$. In the same vein, NV center emission leaking into the SiV⁻ ZPL band as well as rapid hole and electron injection from NVs during charge state initialization further complicate measurements[22,25].

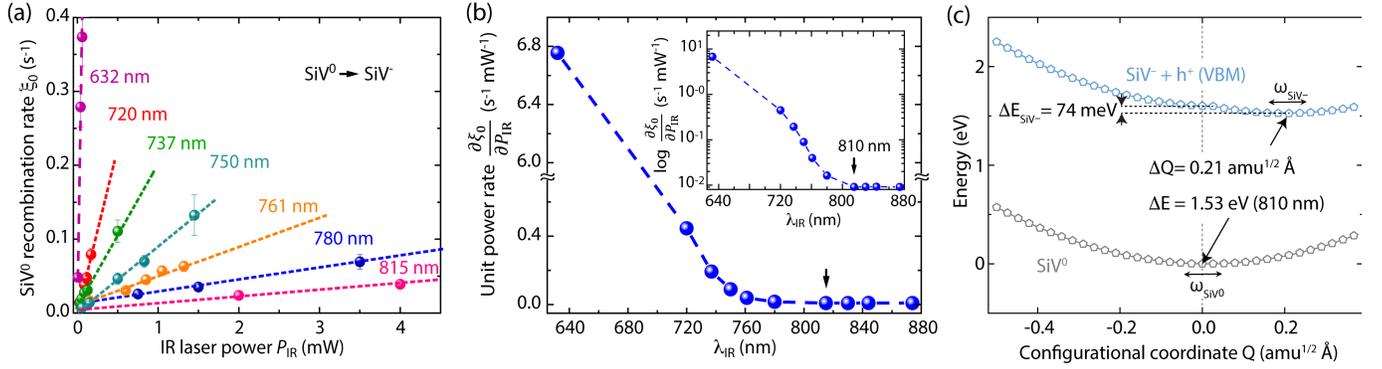

**Figure 2: Recombination of SiV⁰.** (a) SiV⁰ recombination rate, $\xi_0$, as a function of the IR laser power for variable wavelength. Dashed lines are linear fits. (b) Unit power SiV⁰ recombination rates as derived from the linear fits in (a) as a function of wavelength. From the logarithmic plot (upper right insert), we determine the activation energy to be 1.53(5) eV (black arrow). (c) Configurational coordinate diagram for SiV⁻ + h⁺ recombination into SiV⁰. The value of 1.53 eV corresponds to the estimated ionization threshold, while 74 meV is the energy dissipated due to SiV⁻ relaxation.

Fig. 1a schematically lays out our experimental protocol: We first scan a 532 nm beam across an (80 μm)² plane with the laser power (5 mW) and dwell time (1 ms) chosen so as to attain optimal SiV⁻ formation[22]; the plane lies about 5–10 μm below the diamond surface, which eliminates the possibility of surface-induced charge state effects. To produce an SiV⁰-rich area within the plane, we subsequently park the green laser at the center of the scanning range for a variable interval $t_G$. Charge cycling of coexisting NV centers during this time results in the generation of free electrons and holes diffusing away from the illumination region. Preferential hole capture gradually transforms SiV⁻ — as well as any remaining SiV²⁻ — into SiV⁰. Note that the pattern (detected after a 50 μW, 632 nm scan) appears in the form of a "dark halo" since SiV⁰ photoluminescence (centered at 946 nm) is too weak at room temperature.

To characterize SiV⁰ recombination, we park the IR laser at a point within the SiV⁰ pattern (marked with an "X" in the left schematic of Fig. 1b) for a time $t_{IR}$. For a given IR wavelength and power (respectively denoted as $\lambda_{IR}$ and $P_{IR}$ in Fig. 1b), we observe a gradual local conversion of SiV⁰ into SiV⁻, here manifesting through the appearance of a bright spot in the otherwise dark SiV⁰ region (see Supplementary Material, Section 2 for images at all other wavelengths). For the case at hand ($\lambda_{IR} = 720$ nm), Fig. 1c illustrates the measured fluorescence change as a function of $t_{IR}$ for variable IR laser power. To compensate for a change in the halo fluorescence (increasing at longer park time due to scattering of the IR beam), we extract every data point in the plot from the difference between the SiV⁻ fluorescence at site X and the fluorescence from another reference site within the halo not directly exposed to IR. We see in all cases an exponential behavior whose dependence with laser power and wavelength provides the key clues to interpret the underlying recombination mechanism, as we show in the next section.

Conveniently, we can use the same experimental protocol to probe photo-induced SiV⁻ recombination into SiV²⁻ except that in this case we park the IR beam at point "Y" in the SiV⁻-rich section of the plane (see left-hand side schematics in Fig. 1b). Unexpectedly and in contrast to the previous case, infra-red excitation leads here to a reduction of SiV⁻ fluorescence, revealed in Fig. 1b by the appearance of a beam-sized, dark spot in the otherwise bright SiV⁻ region. We capture the recombination of SiV⁻ under 720 nm excitation in Fig. 1d (where, as before, we correct for a change in the starting brightness by subtracting a suitable unaffected reference). The observed SiV⁻ bleaching — evolving non-exponentially to reach a fractional level that depends on the applied laser power (and wavelength) — is somewhat surprising, as the IR photon energy ($< 1.7$ eV) is well below the charge transition threshold to SiV²⁻ (2.1 eV)[26,27], an intriguing response we address below in Section 4 with the help of numerical and *ab initio* modeling.

## 3. Recombination dynamics of SiV⁰

Analysis of the observed SiV⁰ fluorescence with applied IR excitation (Figs. 1c above, see also Supplementary Material, Section 2) allows us to identify the underlying recombination mechanism. Fig. 2a shows the measured recombination rate — extracted from a single exponential fit — as a function of laser power for each of the wavelengths studied in this work. We observe in all cases a linear dependence on light intensity, characteristic of a single photon process. Using the slopes in each data set to calculate the rate coefficients per unit laser power as a function of $\lambda_{\text{IR}}$, we determine the threshold recombination energy as 1.53(5) eV (black arrow in Fig. 2b). We can qualitatively interpret this process as the optical injection of an electron from the valence band into the excited state of SiV⁰ followed by relaxation into the SiV⁻ ground state.

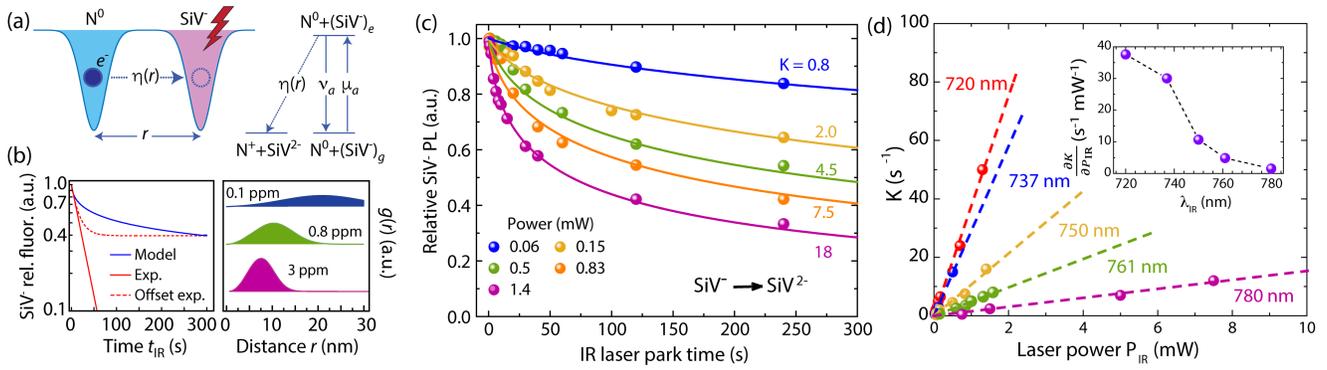

**Figure 3: Nitrogen-assisted recombination of SiV⁻.** (a) Schematics of SiV⁻ recombination. Under optical excitation, the donor electron from a proximal nitrogen tunnels into SiV⁻. Since IR illumination does not excite N⁰, the SiV–N pair can be seen as a three-level system undergoing a photo-induced one-directional transformation into N⁺ + SiV²⁻; the notation follows that in the main text. (b) (Left) Modeled SiV⁻ relative fluorescence assuming $K = 40$ s⁻¹; for comparison, the plot includes a purely exponential decay with the same initial slope (solid red trace) and an exponential decay with an offset adapted to match the end value recorded over the measurement window (dashed red trace). (Right) Calculated nearest-neighbor probability distribution for three example nitrogen concentrations; the blue trace on the left plot uses the distribution corresponding to 3 ppm and an effective tunneling radius $r_0 = 1$ nm. (c) Normalized SiV⁻ fluorescence decay as a function of time after a 720-nm laser parking in the SiV⁻-rich region for different laser powers. Solid lines are fits using Eq. (3) in the main text. (d) Effective SiV⁻ recombination rate $K$ as derived from (c) as a function of laser power for variable IR wavelength; dashed lines are linear fits whose slopes, $\frac{\partial K}{\partial P_{\text{IR}}}$, we plot in the insert as a function of $\lambda_{\text{IR}}$.

To validate the dynamics at play, we model SiV⁻ ↔ SiV⁰ recombination using DFT. By obtaining the adiabatic potential energy surface (PES) for this process (Fig. 2c), we find that the ionization threshold for SiV⁰ into SiV⁻ and a hole (h⁺) in the valence band maximum (VBM) matches very well the onset of the experimental curve in Fig. 2b. The atomic relaxation for this process appears to be rather weak, characterized by small reconfiguration and energy changes (respectively, $\Delta Q = 0.21$ amu$^{1/2}$Å, and $\Delta E = 74$ meV), which, in turn, suggest weak phonon interactions.

## 4. Recombination dynamics of SiV⁻

We now turn to studying the observed time evolution of the SiV⁻ fluorescence, decaying under IR illumination to reach intermediate values dependent on both laser power and wavelength. Adding to the markedly non-exponential time response (see below), this incomplete recombination is difficult to rationalize in a framework that only takes SiV⁻ into account, hence suggesting a more complex scenario. One possibility is to consider the impact of coexisting impurities, substitutional nitrogen being the most natural choice. Known to exist both in neutral and positively charged forms (respectively denoted N⁰ and N⁺), electron transfer to and from substitutional nitrogen has already been invoked to understand charge state interconversion of NVs under continuous illumination in N-rich samples[28]. In the same vein, a charge transfer process involving pairs of proximal NVs and nitrogen — transforming as (NV⁻)* + N⁺ ↔ NV⁰ + N⁰ under green light — seems to underlie recent NV spectroscopy data at varying laser powers[29] (the asterisk denotes the first excited state).

Here we posit that SiV⁻ recombination arises from tunneling of an electron between a photo-excited SiV⁻ and a nearby N⁰, without requiring photoionization of the nitrogen (see Fig. 3a). We start by constructing a phenomenological model using rate equations to describe the observed charge dynamics. We consider a silicon-vacancy and a substitutional nitrogen separated by a distance $r$, and assume that only SiV⁻ undergoes optical excitation under IR illumination (from the ground ($g$) to its first excited state ($e$), see Supplementary Material, Section 3). This scenario is the most reasonable as the IR photon energies are below the 2.2 eV threshold where significant N⁰ photoionization onset occurs[30-32]. We note this is significantly above the N⁰ donor level (1.7eV) due to the additional energy required to redistribute the position of N and C atoms following loss of an electron[33]. On the other hand, IR excitation below the SiV–ZPL is facilitated by anti-Stokes absorption[34].

Then, the rate equations that describe the SiV⁻ ($X_{\text{SiV}}^{g,e}$) and SiV²⁻ ($X_{\text{SiV}}^{2-}$) populations in the simultaneous presence of IR excitation and tunneling are given by

$$\frac{dX_{SiV}^g}{dt} = -\mu_{SiV}X_{SiV}^g + \nu_{SiV}X_{SiV}^e, \quad (1a)$$

$$\frac{dX_{SiV}^e}{dt} = \mu_{SiV}X_{SiV}^g - \nu_{SiV}X_{SiV}^e - \eta(r)X_N^0 X_{SiV}^e. \quad (1b)$$

In the above formulas, $\mu_{SiV}$, $\nu_{SiV}$ respectively represent the SiV$^-$ IR excitation and relaxation rates, $X_N^0 = 1 - X_N^+$ denotes the N$^0$ population, and $\eta(r)$ is the probability per unit time characterizing the electron transfer as a function of distance; we assume $\eta(r) = Ce^{-r/r_0}$ where $C$ is a constant and $r_0$ represents an effective tunneling radius[35]. Since, in general, $\mu_{SiV}, \nu_{SiV} \gg \eta$, we can conveniently isolate the dynamics describing tunneling and recast Eqs. (1a) and (1b) under IR illumination as

$$\frac{d(1-X_{SiV}^{2-})}{dt} = -\eta(r)\frac{\mu_{SiV}\xi_N}{\mu_{SiV}+\nu_{SiV}}\left(1-X_{SiV}^{2-}\right)^2, \quad (2)$$

where $1-X_{SiV}^{2-} = X_{SiV}^g + X_{SiV}^e$ represents the SiV$^-$ population at a given time (our observable, see schematics in Fig. 1a). We use in Eq. (2) the expression $X_N^0 = \xi_N(1-X_{SiV}^{2-})$, where $\xi_N = \xi_N(\lambda_{IR}, P_{IR})$ is, in general, a function of laser wavelength and power. This relation effectively ties the N charge state at a given time $t_{IR}$ to that of the proximal SiV and is valid herein assuming the nitrogen concentration is sufficiently low (Supplementary Material, Section 3). Solving Eq. (2), we find

$$\overline{(1-X_{SiV}^{2-})}(t_{IR}) = 4\pi \int_0^\infty \frac{g(r)}{1+K\,t_{IR}\,e^{-r/r_0}} r^2 dr, \quad (3)$$

where the upper bar denotes an average over all SiV–N pair distances, $g(r)$ is a weight given by the nearest-nitrogen probability distribution (itself a function of the nitrogen concentration, see Fig. 3b), and $K = K(\lambda_{IR}, P_{IR}) = \frac{C\,\mu_{SiV}\xi_N}{\mu_{SiV}+\nu_{SiV}}$ is a fitting parameter dependent on the operating laser wavelength and power.

Adding to the results in Fig. 1d, the solid circles and lines in Fig. 3c respectively show the relative SiV$^-$ fluorescence as a function of time for $\lambda_{IR} = 750$ nm and the calculated fits using Eq. (3) assuming $r_0 = 1$ nm, as seen for NV centers[36]; in all cases, we set the nitrogen concentration at 3 ppm, consistent with prior observations in this same sample crystal[22]. Comparison between the measured and calculated SiV response shows good agreement (see also Supplementary Material, Section 2 for the complete data set). Importantly, Eq. (3) captures the non-exponential response with the IR illumination time as well as the fractional SiV$^-$ charge state conversion over the finite duration of the experiment, a combined consequence of the non-linearity of Eq. (2) and the varying contributions from pairs separated by growing distances (see also left plot in Fig. 3b).

Fig. 3d shows the extracted $K$ values as a function of laser power, showing a linear dependence for all observed wavelengths. To interpret this observation, we note that in the low laser intensity limit — where $\nu_{SiV} \gg \mu_{SiV} \propto P_{IR}$, expected here — Eq. (3) predicts $K \propto \xi_N\mu_{SiV}$, which implies a linear dependence with laser power provided $\xi_N$ is insensitive to the excitation intensity.

The insert in Fig. 3d shows the extracted slope, $\frac{\partial K}{\partial P_{IR}}$, as a function of $\lambda_{IR}$; we find the process activates near the SiV$^-$ ZPL, consistent with a charge transfer model where SiV$^-$ excitation is a necessary condition[34]. These observations alone, however, are insufficient to validate the energetics at play or understand the microscopic mechanisms driving electron tunneling from N$^0$; we tackle these key aspects immediately below with the help of DFT.

## 5. Atomistic modeling of electron transfer

We carry out all DFT calculations in a 512-atom supercell via the VASP software package[37] and using either the PBE[38] or HSE06[39] exchange-correlation functionals (Supplementary Material, Section 5). We begin by determining the charge state transition energies for SiV and substitutional N (Fig. 4a): In agreement with prior work[27,28,40-42], the states most likely involved in an N-mediated recombination of SiV$^-$ (other than the starting SiV$^-$ and N$^0$) are SiV$^{2-}$ and N$^+$, which we now make the focus of our analysis. From the DFT energy levels for these defect charge states (Fig. 4b), it is worth highlighting a few key features: (*i*) N$^+$ has no occupied intra-bandgap defect states, with its lowest-unoccupied molecular orbital (LUMO) lying close to the conduction band minimum (CBM); (*ii*) besides the "donor electron" state — here corresponding to the highest occupied molecular orbital (HOMO) — N$^0$ also features two electronic states below the valence band maximum (VBM) whose wavefunctions show small but non-negligible defect-like contributions, which we marke as "HOMO-1" and "HOMO-2", (*iii*) the $e_u$ state of SiV$^-$ also lies within the valence band, and appears to have an energy slightly lower than that of the mixed HOMO-1 state in N$^0$; (*iv*) photo-excitation of SiV$^-$ propels an $e_u$ electron of the spin minority channel into the LUMO state ($e_g$); (*v*) all relevant defect states in SiV$^{2-}$ are occupied.

Building on the above observations, the charge transfer mechanism we propose starts with the optical excitation of SiV$^-$ into (SiV$^-$)$^*$, a process that leaves behind an unoccupied $e_u$ state (stage ① in Fig. 4c). The crucial step in the charge transfer is carrier tunneling, occurring through the relocation of a HOMO-1 or HOMO-2 electron into the (now vacant) SiV $e_u$ orbital (stage ②); relaxation of the nitrogen impurity — transiently left in an excited state we refer to as (N$^+$)$^{**}$ — drives N$^+$ into the ground state irreversibly, in the process trapping the system into the $N^+ + SiV^{2-}$ state (stage ③).

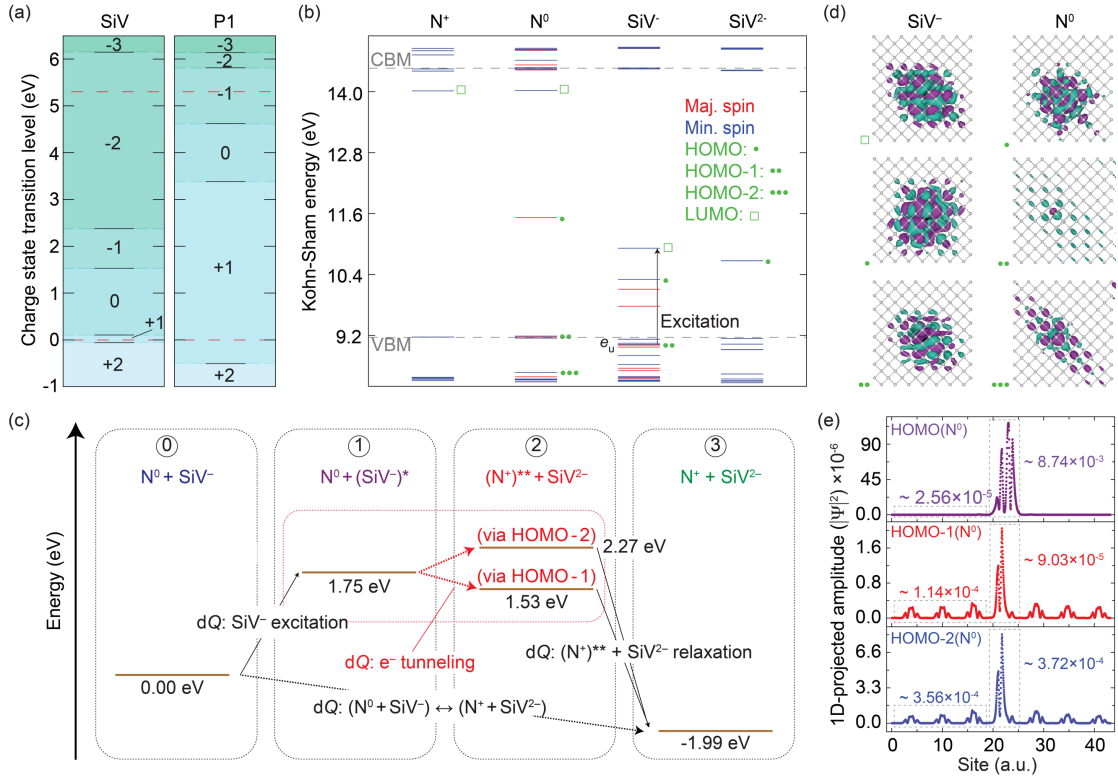

**Figure 4.** *Ab initio* modelling of N-mediated SiV⁻ recombination. (a) Thermodynamic charge-state transition levels for (left) SiV⁻ and (right) N centers. (b) Spin-resolved electronic levels of N and SiV centers for the charge states relevant to our model. We use a square and an asterisk to respectively highlight the highest-occupied and lowest-unoccupied states (HOMO and LUMO, respectively); multiple asterisks indicate other defect (or defect-mixed) states, ordered by the energy difference with respect to their HOMO. (c) Energy-level diagram as derived from PES minima for different SiV/N excitation, ionization, and recombination pathways. All levels are given with respect to the $N^0 + SiV^-$ configuration, which is taken as the initial state for the whole charge transfer process. For reference, the diagram also includes the nature of the coordinate reconfigurations along each step. (d) Real part of the wavefunction for the states highlighted in (b), see side labels. Here, cyan is positive and purple is negative; black, red, and gray spheres represent Si, N and C atoms, respectively. (e) Line-cuts along the ⟨111⟩ axis of the wavefunction magnitude for the three highest-energy occupied states of $N^0$. The integrated values along this direction are included for comparison. The supercell size is 512 atoms in (a) and (b) and 2474 atoms in (d) and (e); all calculations are carried out at the HSE06 level.

A key feature in this mechanism is the nature of the states involved: Unlike the $N^0$ donor electron state, the HOMO-1 and HOMO-2 orbitals are comparatively delocalized due to hybridization with valence band states, arguably facilitating electron tunneling from $N^0$ to $(SiV^-)^*$. We gauge the spatial spread in Fig. 4d, where we boost the size of the supercell from 512 atoms — used in all other calculations — to 2474 atoms, and determine the wavefunctions of all relevant orbitals. We compare line cuts for the $N^0$ wavefunction magnitudes along the ⟨111⟩ axis in Fig. 4e, and find that the two sub-VBM orbitals feature non-negligible amplitudes across the supercell. Importantly, the electron densities we calculate for the HOMO-1 and HOMO-2 orbitals in areas farther removed from the defect site are orders of magnitude larger than those of the $N^0$ donor level, hence facilitating the charge transfer. On a related note, we also highlight the close alignment between the energies corresponding to the $N^+ + (SiV^-)^*$ and $(N^+)^{**} + SiV^{2-}$ transients (see stages ① and ② in Fig. 4c), which makes this type of transfer close to resonant.

Aside from the energetics and spatial characteristics of the electronic wave functions, one last factor of interest is the level of atomic reconfiguration in the charge transfer process. By looking at the adiabatic PES of alternative $N^0$ ionization pathways, we find a stark contrast between the extent of atomic reconfiguration when ionized via its HOMO state or via "HOMO-1" or "HOMO-2", where the net configurational displacement (d$Q$) is about 20-fold smaller (between 0.05–0.06 amu$^{1/2}$ Å), which facilitates the charge transfer. These latter displacements are also akin to those obtained for processes involving the SiV⁻, which for the most part remain relatively small (< 0.25 amu$^{1/2}$ Å, see also Supplementary Material, Sections 4 through 6).

## 6. Discussion

In summary, we presented a comprehensive data set on the photo-induced charge state dynamics of SiV throughout the important energy range where both $SiV^0$ and $SiV^-$ recombination activate. Our experiments capitalize on the ability to initialize the SiV charge on demand, either by direct optical excitation or through capture of carriers photo-generated non-locally. Building on *ab initio* modeling, we gained a microscopic understanding and, with it, a unified view of the charge state dynamics at play. Specifically, we can associate $SiV^0$ recombination to a one-photon process involving the injection of an electron from the valence band. In a similar manner, our experimental and theoretical work on $SiV^-$ recombination supports the notion of a photo-activated electron transfer from a proximal nitrogen donor. Rather than the $N^0$ donor electron, carrier transfer takes place from intra-valence-band orbitals whose energies are nearly aligned with that of the $SiV^-$ electron undergoing optical excitation. Combined, these results are consistent with photoconductivity measurements of bulk samples[43,44], recent observations of bound excitonic states[17] in $SiV^0$, as well as preceding *ab initio* calculations on SiV[26,27].

Given the growing interest on the use of SiV centers for quantum information processing and nanoscale sensing applications, our findings have implications for the design and operation of devices that capitalize on the SiV physical properties. For example, the charge transfer processes unraveled here could serve as integral precursor steps to implementing optimized photoelectric detection of silicon vacancy centers, and in turn, understanding the generation of photocurrent in diamonds containing unwanted silicon impurities[45]. In the same vein, manipulation of $SiV^-$ as a stable photon source will benefit from additional efforts in reducing the concentration of proximal donors. Interestingly, a strong suppression of blinking and spectral diffusion in single $SiV^-$ has been recently seen upon prolonged exposure to blue light[46], suggesting that local ionization of $N^0$ into $N^+$ could be responsible for the observed charge state stabilization.

While our modeling is consistent with nitrogen impurities serving as the electron source for $SiV^-$ recombination, additional work — e.g., involving the systematic characterization of diamond samples with variable nitrogen content — will be needed to conclusively assign the nature of the donor point defect. The limit of higher nitrogen concentration is particularly interesting as it is likely to alter the condition $\xi_b \sim 1$ valid herein, and correspondingly change the observed fluorescence time response with laser power and wavelength.

Different charge initialization protocols — relying on green laser scans or capture of carriers diffusing from a non-local green park — can arguably lead to nitrogen ensembles with distinct charge state compositions. Future theoretical and experimental work should therefore address the impact of these changes on the SiV recombination dynamics, specifically concerning the ability to augment or suppress charge tunneling from N to SiV. Similarly, it will be important to contemplate charge states in nitrogen beyond $N^0$ and $N^+$ considered here. Of particular interest is $N^-$, proposed as a transient charge state to interpret absorption spectroscopy results[47] but possibly present as a metastable, long-lived charge configuration in crystals with sufficiently low nitrogen concentration[48].

Although the present experiments are limited to ambient conditions, similar responses are also likely at lower temperatures as the invoked charge transfer mechanisms — namely, one-photon optical injection and electron tunneling — are expected to be largely temperature insensitive. Something equivalent can be said about extensions from SiV ensembles to individual color centers, provided the nitrogen concentration remains moderately high. Lastly, the ability to initialize SiV into the neutral state combined with tunable laser excitation slightly below the ionization threshold open opportunities to further examine the formation of bound excitonic states in the absence of detrimental dopants[17] (previously required to stabilize $SiV^0$).

**Acknowledgements**

We thank Artur Lozovoi for helpful discussions. G.G.A. and C.A.M. acknowledge funding by the U.S. Department of Energy, Office of Science, National Quantum Information Science Research Centers, Co-design Center for Quantum Advantage (C2QA) under contract number DE-SC0012704. J.F. acknowledges support from the National Science Foundation through grant NSF-2216838; G.L.M. acknowledges funding from grant NSF-2208863. A.A.W. was supported by the Australian Research Council grant ID DE210101093, and N.B.M. acknowledges support from the Australian Research Council, DP1700100169. All authors also acknowledge access to the facilities and research infrastructure of the NSF CREST IDEALS, grant number NSF-2112550. The Flatiron Institute is a division of the Simons Foundation.

**Conflict of Interest**

The authors declare no conflict of interest.

**Data Availability Statement**

The data that support the findings of this study are available from the corresponding author upon reasonable request.

SUPPLEMENTARY MATERIAL

# Photo-induced ionization dynamics of the neutral and negatively charged silicon vacancy defect in CVD diamond at room temperature

G. Garcia-Arellano[1], G. López-Morales[1], N. B. Manson[4], J. Flick[1,2,3], A. A. Wood[5], and C. A. Meriles[1,2]

[1]Department of Physics, CUNY-City College of New York, New York, New York 10031, USA.

[2]CUNY-Graduate Center, New York, NY 10016, USA.

[3]Center for Computational Quantum Physics, Flatiron Institute, New York, NY 10010, USA.

[4]Department of Quantum Science and Technology, Research School of Physics, Australian National University, Canberra, A.C.T. 2601, Australia

[5]School of Physics, The University of Melbourne, Parkville VIC 3010 Australia

1. **Experimental methods: Multicolor confocal microscopy at room temperature.**
2. **$SiV^0$ and $SiV^-$ ionization dynamics under near-IR excitation.**
3. **Nitrogen-assisted electron tunneling.**
4. **$SiV^0$ and $SiV^-$ ionization dynamics in a different CVD sample.**
5. **Computational methods.**
6. **Configurational coordinate diagrams (CCDs) and electron transfer via $N^0$'s highest occupied molecular orbital.**

## 1. Experimental methods: Multicolor confocal microscopy at room temperature

The experimental setup consists of a multi-color confocal microscope with two branches for visible and the infrared excitation (Figure S1). For the visible branch, we use a 40-mW laser at 532 nm sourced from a DPSS laser diode module (Thorlabs DJ532-40) and a 633-nm red light from a 70-mW laser diode (Thorlabs HL63163DG). Light from both sources is combined via a dichroic mirror (Thorlabs DMLP638) and coupled into a single-mode fiber. The output from the fiber is collimated with an NA = 0.5 achromatic objective lens (Olympus UMPlanFl N). The scanning confocal microscope uses a two-axis galvo steering mirror system (Thorlabs GVS002), a 4$f$-relay lens ($f_1$ = 100 mm, $f_2$ = 200 mm), and a NA = 0.4 microscope objective (Mitutoyo MPlan Apo 2) mounted on a one-axis piezo scanning stage [1]

For the near-infrared excitation, we use a Ti:Sapphire laser (Coherent Mira 900) tunable between 700 and 900 nm and run in cw-mode. The output beam of the laser is intensity controlled with an acousto-optic modulator (AOM, Isomet 1250 C-829A) and thereafter sent into a single-mode optical fiber. The output from the fiber, after collimation, is reflected off a motorized flip mirror on the transmission side of the dichroic mirror, that allows to switch between the near-IR excitation and readout. We control the beam intensity using a variable neutral density filter at the output of the fiber.

The light from the diamond is collected using the reverse optical path and subsequently separated from the excitation light by the dichroic mirror. To select regions of the PL spectrum we first use 700 nm and 650 nm long-pass filters to prevent red light leakage to the detection, and a 735-nm, 10-nm-wide band pass filter to select out the SiV$^-$ zero-phonon line (ZPL). The PL is then focused using a 12.5-mm aspheric lens into a single-mode fiber (MFD = 10 µm) and directed into a single-photon counting module (SPCM, Excelitas SPCM-AQRH-14). Photon counts are recorded using a National Instruments NI-PCIe 6321 data acquisition card, which additionally supplies the voltage for driving the galvo scanning system, the AOM and motorized flip mirror.

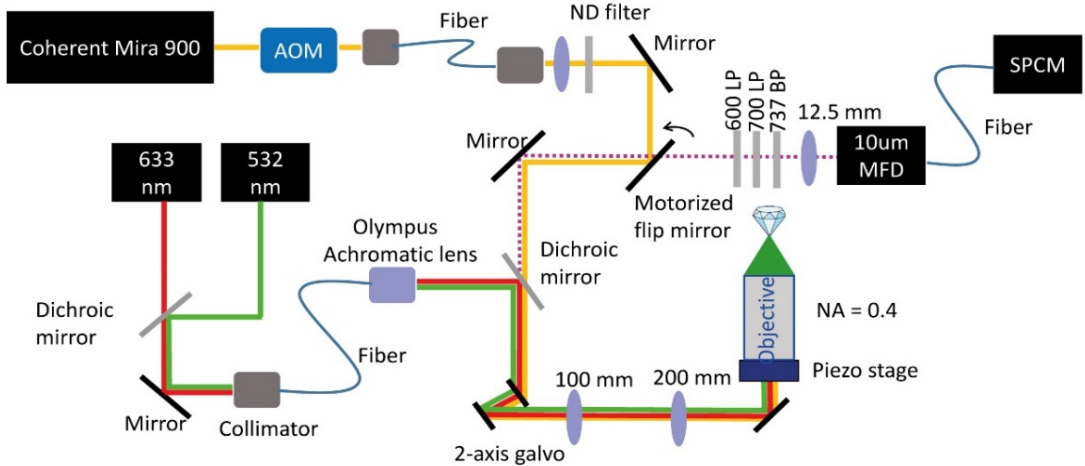

**Figure S1.** Multicolor confocal microscope used to study the recombination dynamics of SiV$^-$ and SiV$^0$ under near-IR illumination.

## 2. SiV$^0$ and SiV$^-$ ionization dynamics under near-IR illumination

In Figures S2-S6 (a) we present examples of confocal images obtained to study the ionization dynamics of the neutral (SiV$^0$) and negatively (SiV$^-$) charge states of the silicon vacancy defect at different wavelengths in the range 720-780 nm, under the protocol described in Fig. 1(a) of the main text. The IR laser is parked at the SiV$^0$ (X point) and SiV$^-$ (Y point) as indicated in the diagram of Fig. S2(a) for a variable park time $t_{IR}$. The wavelength, power and parking time are indicated in each of the figures.

Figures S2-S6 (b) present the integrated fluorescence (solid points) vs parking time obtained from the excitation of the photo-generated SiV$^0$ under near-IR illumination. The solid lines are exponential fits to the data. From each fit we extract the SiV$^0$ ionization rate ($\xi_0$) at a given power and we plot them in Fig. 2(a) of the main text. For all wavelengths, we observe a linear dependence with the power, a hallmark of a single photon process. To characterize the linear behavior, we have also plotted in Fig. (2b) of the main text the slope (or unit power rate $\frac{\partial \xi_0}{\partial P_{IR}}$) as a function of laser wavelength.

Similarly, Figures S2-S6 (c) show the integrated fluorescence data (solid points) vs parking time obtained from the excitation of SiV$^-$ with near IR illumination. At all wavelengths in the range 720 - 780 nm we observe a non-exponential decay of the fluorescence. The solid lines are fits based on the nitrogen-assisted model described below (see section 3). The fitting parameter $K$ extracted at each power is reported in Fig. 3(d) of the main text.

In order to determine the SiV$^0$ recombination threshold with a better accuracy, we performed additional measurements at wavelengths longer than 780 nm. Fig. S7 shows the integrated fluorescence (solid points) vs parking time, obtained when parking on the SiV$^0$ region (X point) with a) 633 b) 815 and c) 830, 844 and 874 nm illumination. The SiV$^0$ unit power rates for each wavelength are presented in Fig. 2(b) of the main text.

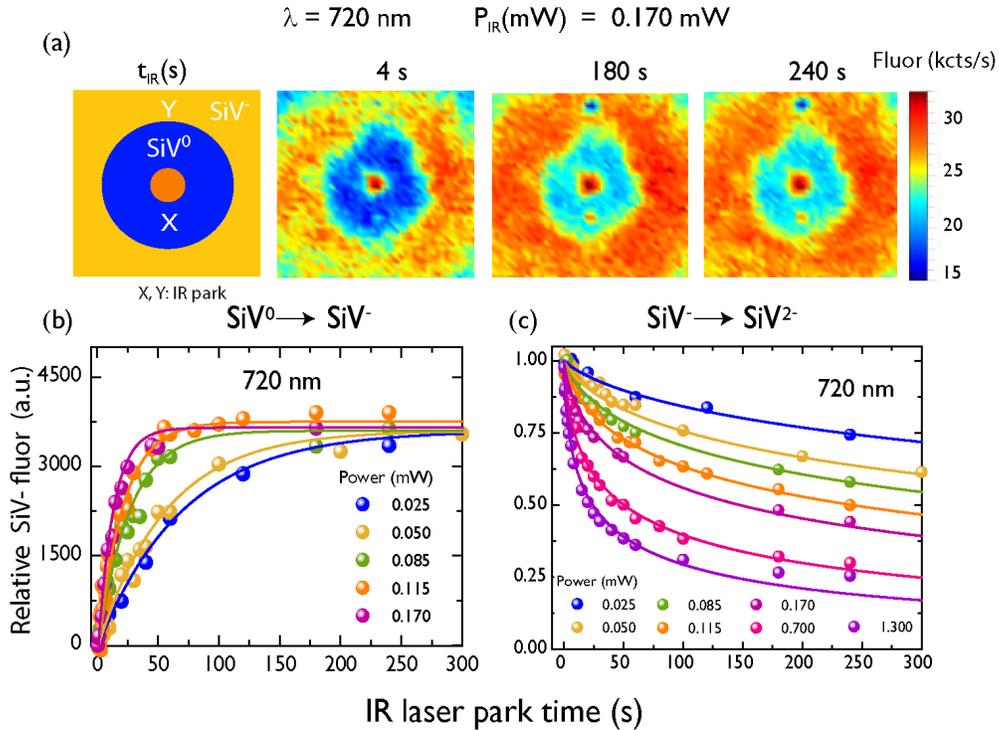

**Figure S2:** See captions below.

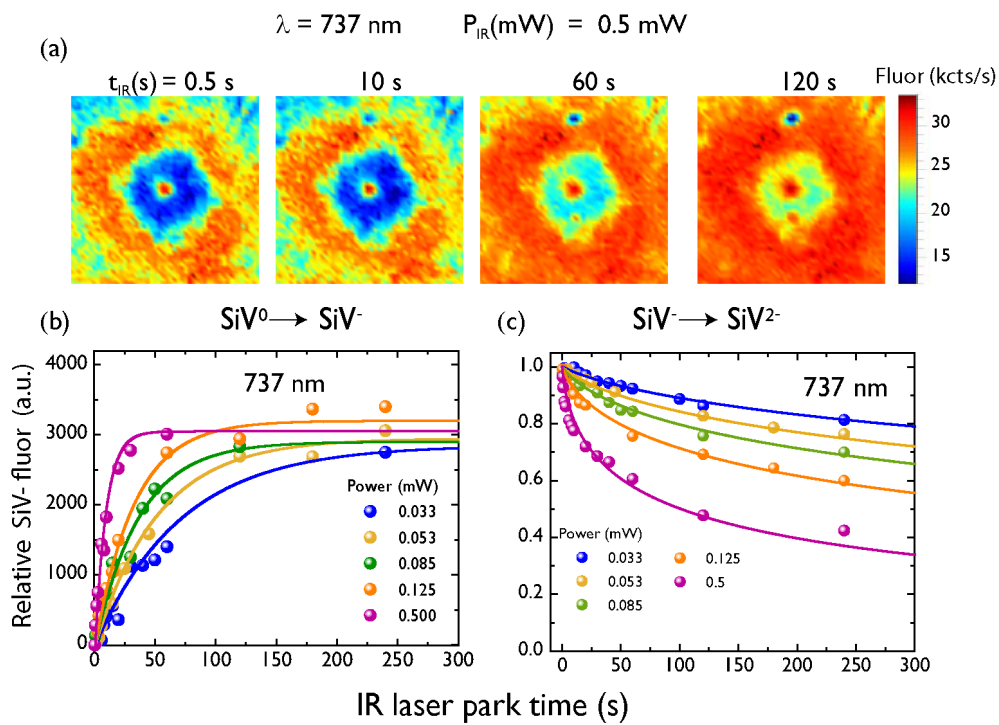

**Figure S3:** See captions below.

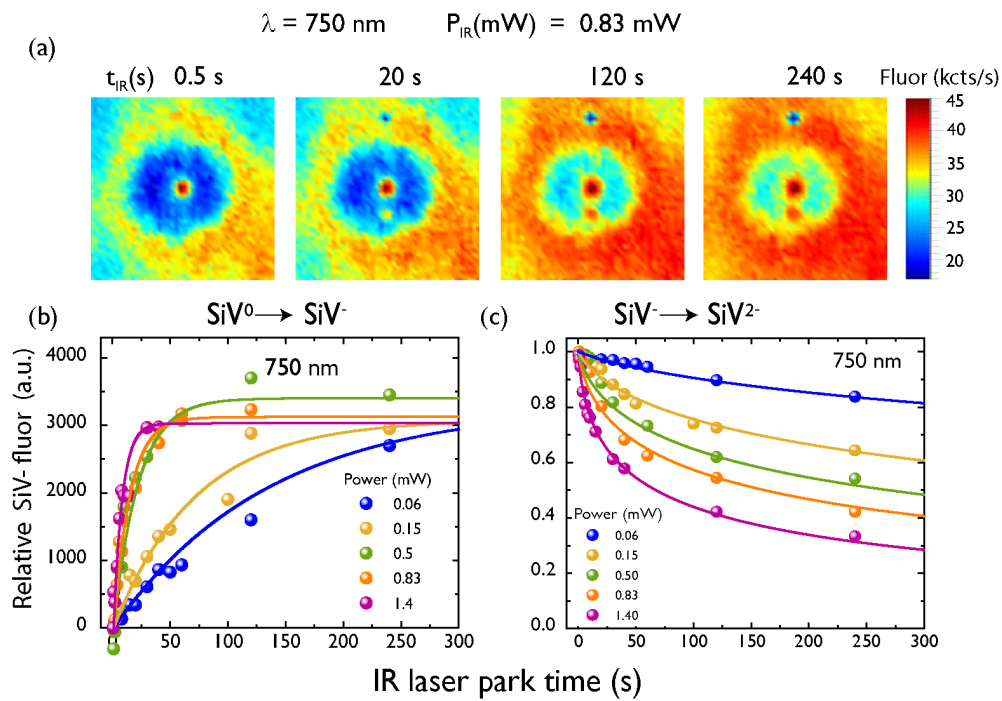

**Figure S4:** See captions below.

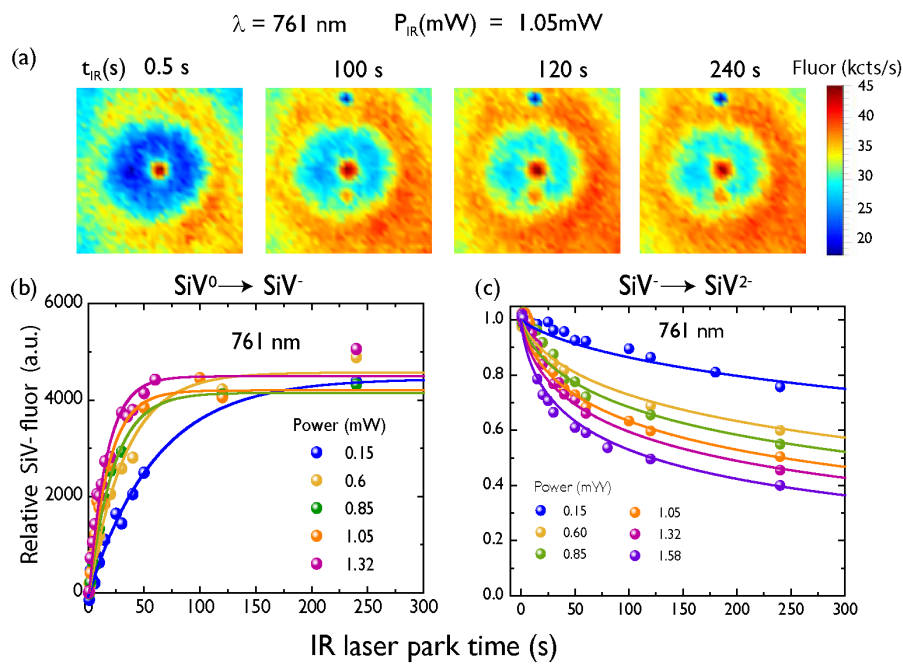

**Figure S5:** See captions below.

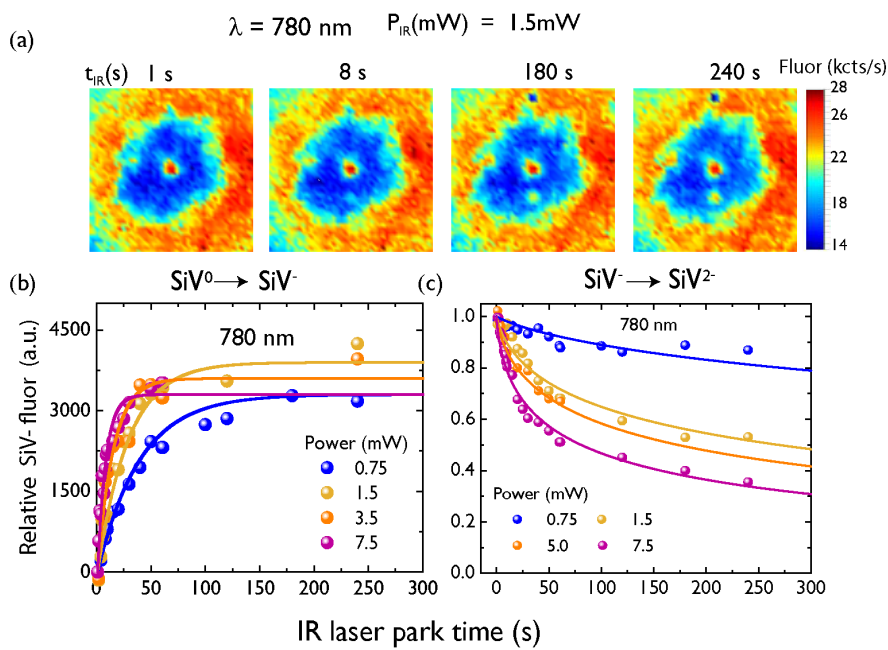

**Figure S6:** See captions below.

**Figures S2- S6:** (a) Confocal images obtained following the protocol described in Fig 1(a) of the main text. After photogeneration of the SiV$^0$ pattern, the IR laser is parked at the position X (SiV$^0$ region) and Y (SiV$^-$ region) at different powers and wavelengths as indicated in the diagram of figure S2(a). (b) Integrated SiV$^-$ fluorescence (solid points) versus parking time obtained at different powers of the Ti:Sa laser at the wavelength in the SiV$^0$ region (X point). The solid lines are exponential fits to the data The rate extracted from the fits at each wavelength is reported in Fig. 2(a) of the main text). (c) Integrated SiV$^-$fluorescence (solid points) versus parking time obtained at the SiV$^-$ region. The solid lines are fits to the data using the nitrogen assisted electron tunneling model. (see section 3 below). The coefficient K extracted at each fit is reported in Fig. 3(d) of the main text.

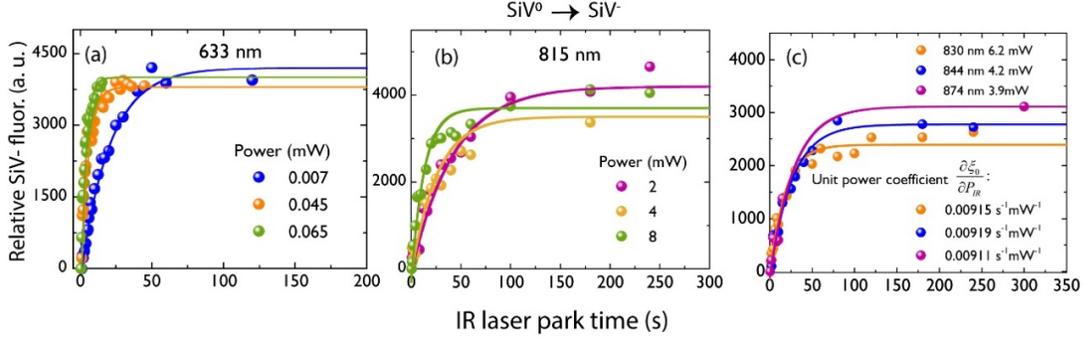

**Figure S7:** Integrated fluorescence (solid points) vs parking time obtained from the excitation of $SiV^0$ (X point) with a) 633nm, b) 815 nm and c) 830, 844, 874 nm illumination. The $SiV^0$ unit power rates are reported in Fig. 2(b) of the main text.

## 3. Nitrogen-assisted electron tunneling

We focus our attention now on modelling the recombination process of $SiV^-$ into $SiV^{2-}$ under near-IR illumination. We start by considering a defect pair comprising a negatively charged SiV and a neutral N separated by a distance $r$, and posit that $SiV^-$ recombination arises from tunneling of the donor electron from $N^0$ during photoexcitation of $SiV^-$. Despite its nominally low ionization energy (~1.7 eV or, equivalently, 730 nm), we assume $N^0$ does not undergo excitation since its absorption cross section is low in the wavelength range relevant to these experiments (see Fig. 3(a) of the main text). The rate equations governing the dynamics can then be cast as

$$\frac{dX_{SiV}^g}{dt} = -\mu_{SiV} X_{SiV}^g + \nu_{SiV} X_{SiV}^e \tag{S1a}$$

$$\frac{dX_{SiV}^e}{dt} = \mu_{SiV} X_{SiV}^g - \nu_{SiV} X_{SiV}^e - \eta(r) X_N^0(\vec{r}, t) X_{SiV}^e(0, t) \tag{S1b}$$

where $X_{SiV}^g$ ($X_{SiV}^e$) represents the fractional $SiV^-$ population in the ground (excited) state, $\mu_{SiV}$ and $\nu_{SiV}$ respectively denote the $SiV^-$ optical excitation and relaxation rates, and $\eta(r)$ represents the unit time probability characterizing the electron transfer. Note that the probability of finding the proximal nitrogen in the neutral charge state, $X_N^0(\vec{r}, t)$, is, in general, a function of time $t$ and its position $\vec{r}$ relative to the SiV (assumed at the origin). On the other hand, the SiV charge states satisfy the conservation equation $X_{SiV}^g + X_{SiV}^e + X_{SiV}^{2-} = 1$ with $X_{SiV}^{2-}$ denoting the fractional population in the $SiV^{2-}$ charge state. Then, we recast Eq. S1(a) as

$$\frac{dX_{SiV}^g}{dt} = -(\mu_{SiV} + \nu_{SiV}) X_{SiV}^g + \nu_{SiV}(1 - X_{SiV}^{2-}) \tag{S2}$$

Assuming that $X_{SiV}^{2-}$ changes very slowly in comparison with the excitation and relaxation rates $\mu_a$, $\nu_a$, then the solution to Eq. S2 takes the form $X_{SiV}^g = A + Be^{-\kappa t}$ with $A$, $B$, and $\kappa$ given by $A = \frac{\nu_{SiV}}{\mu_{SiV}+\nu_{SiV}}(1 - X_{SiV}^{2-})$, $B = \frac{\mu_{SiV}}{\mu_{SiV}+\nu_{SiV}}(1 - X_{SiV}^{2-})$, and $\kappa = \mu_{SiV} + \nu_{SiV}$

We can now rewrite $X_{SiV}^g$ as

$$X_{SiV}^g = (1 - X_{SiV}^{2-})\frac{1}{\mu_{SiV}+\nu_{SiV}}\left(\nu_{SiV} + \mu_{SiV} e^{-(\mu_{SiV}+\nu_{SiV})t}\right). \tag{S3}$$

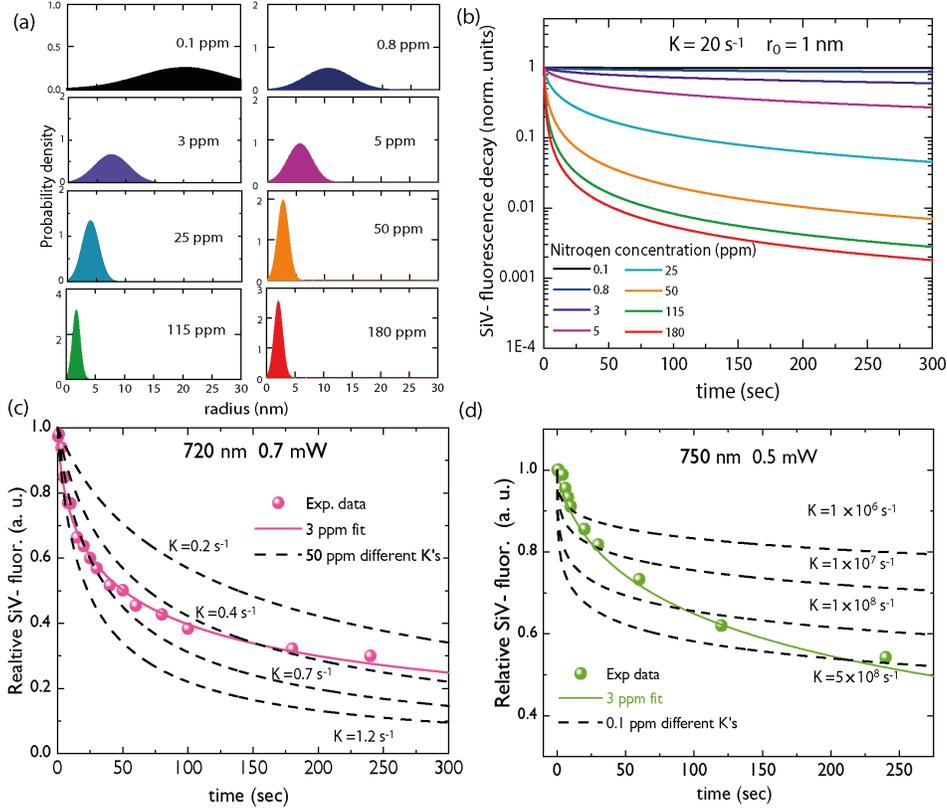

**Figure S8:** (a) Nearest-neighbor probability distribution for different nitrogen concentrations. (b) Calculated SiV⁻ fluorescence obtained from Eq. S8 assuming that the initial SiV⁻ concentration is fixed and illuminated under near-IR at fixed power for different nitrogen concentrations. (d) Comparison of the data at 720 nm and 0.7 mW (solid points) fitted using a nearest-nitrogen distribution corresponding to 3 ppm (solid pink line) and the calculated SiV⁻ dynamics with a nitrogen distribution of 50 ppm for different $K$ values. (c) Comparison between the data at 750 nm and 0.5 mW (solid points) fitted using a nearest-nitrogen probability distribution corresponding to 3 ppm (solid green line) and the expected SiV⁻ dynamics with a nitrogen distribution of 0.1 ppm for varying $K$. The tunneling radius is fixed at $r_0 = 1$ nm.

Similarly, we express $X_{\text{SiV}}^e$ as

$$X_{\text{SiV}}^e = \left(1 - X_{\text{SiV}}^{2-}\right) - X_{\text{SiV}}^g$$

$$= \left(1 - X_{\text{SiV}}^{2-}\right)\frac{\mu_{\text{SiV}}}{\mu_{\text{SiV}}+\nu_{\text{SiV}}}\left(1 - e^{-(\mu_{\text{SiV}}+\nu_{\text{SiV}})t}\right) \quad \text{(S4)}$$

Adding Eqs. S1(a) and S1(b), we then obtain

$$\frac{d(1-X_{\text{SiV}}^{2-})}{dt} = -\eta(r)X_{\text{SiV}}^e(\vec{r},t)X_N^0(\vec{r},t) = -\eta(r)\frac{\mu_{\text{SiV}}}{\mu_{\text{SiV}}+\nu_{\text{SiV}}}\left(1 - X_{\text{SiV}}^{2-}\right)X_N^0(\vec{r},t), \quad \text{(S5)}$$

where we assumed $X_{\text{SiV}}^e$ (and $X_{\text{SiV}}^g$) take their time-independent limit values throughout the evolution, an approximation justified given that $(\mu_{\text{SiV}} + \nu_{\text{SiV}})^{-1}$ is a very short time scale compared to that defined by the inverse tunneling rate, $\eta^{-1}$. For excitation wavelengths above 720 nm, no ionization of N⁰ has been reported, hence allowing us to write $X_N^0(\vec{r},t) = \xi_N\left(1 - X_{\text{SiV}}^{2-}\right)$, where $\xi_N =$ $\xi_N(\lambda_{IR}, P_{IR})$ is, in general, a function of laser wavelength and power. Here we assume that $\eta(r)$ decreases exponentially with the distance between the SiV⁻ and N⁰, and is given by the exponential function $\eta(r) = Ce^{-r/r_0}$ with $r_0$ denoting the effective tunneling radius. The solution to Eq. (S5) is then

$$(1 - X_{SiV}^{2-}(t)) = \frac{1}{1+K\ t_{IR}\ e^{-r/r_0}}, \quad (S6)$$

where $K(\lambda_{IR}, P_{IR}) = \frac{C\ \mu_{SiV}\xi_N}{\mu_{SiV}+\nu_{SiV}}$ is a fitting parameter dependent on the operating laser wavelength and power. Finally, if we consider an ensemble of pairs separated by a variable distance $r$, the fluorescence decay — governed by $(1 - X_{SiV}^{2-}(t))$ — is given by the expression

$$\overline{(1 - X_{SiV}^-)} = 4\pi \int_0^\infty r^2 \frac{g(r)}{1+K\ t_{IR}\ e^{-r/r_0}} dr, \quad (S7)$$

where the upper bar denotes an average over all SiV–N pair distances, and $g(r)$ represents the nearest neighbor probability distribution, in turn a function of the nitrogen concentration (Fig. S8(a)) [3]. In Fig. S8(b) we plot the expected SiV⁻ fluorescence decay at fixed power ($K = 20$ s⁻¹) and radius $r_0 = 1$ nm obtained from Eq. S7. As we can see, if the nitrogen concentration increases, a larger fraction of the SiV⁻ initial population is transformed into SiV²⁻ via electron tunneling.

To analyze the data shown in Figs. S2(c) to S6(c) we set the nitrogen concentration to 3 ppm, consistent with prior observations in this sample [2] and $\xi_b \sim 1$. As shown in Fig. 3(d) of the main text, the fitting parameter $K$ exhibits a linear dependence with the power for all wavelengths in the range 720-780 nm. Consistent with this observation, we expect $\nu_{SiV} \gg \mu_{SiV} \propto P_{IR}$ in the limit of low laser intensities, implying that $K \propto \xi_N\mu_{SiV}$ and thus linear with laser power (provided $\xi_N$ insensitive to the excitation intensity).

In Figs. S8(c) and S8(d) we show the expected SiV⁻ fluorescence decay calculated using nitrogen concentrations one order of magnitude smaller (S8(c): 0.1 ppm) and one order of magnitude larger (S8(d): 50 ppm) than the one estimated in the sample (solid lines). As we can see nitrogen concentrations different to the 3-5 ppm expected for this diamond crystal do not properly reproduce the experimental data.

## 4. SiV⁰ and SiV⁻ dynamics in a different CVD sample

In this section we present ionization measurements for a different CVD sample (named sample B). The SiV and NV concentrations are 0.9 ppm and 10 ppb respectively [3]. Figure S9(a) shows the confocal images obtained under the protocol presented in the main text (Fig. 1(a)). The IR laser (750 nm) is parked in the SiV⁰ region (X point) and SiV⁻ region (Y point) at different powers. Figure S9(b) shows the integrated fluorescence obtained after illumination of the SiV⁰ region at different powers. The SiV⁰ ionization rate at each power is presented in Fig. S6(d). The extracted unit power rate $\xi_0$ matches with the one found in the sample presented in the main text (see blue curve at 750 nm in Fig. 2(a) of the main text).

Figure S9(c) presents the integrated SiV⁻ fluorescence (solid circles) vs time at point Y. In agreement with the sample studied in the main text, we also observe a non-exponential decay of the SiV⁻ fluorescence under near-IR illumination. The solid lines represent fits to the model of nitrogen-assisted electron tunneling (see section 3) using a nitrogen concentration of 3 ppm. The coefficient $K$ versus power obtained from the fit is presented in Fig. S9(e). The slope that characterizes the linear dependence of $K$ with power is comparable to the one obtained in the sample studied in the main text.

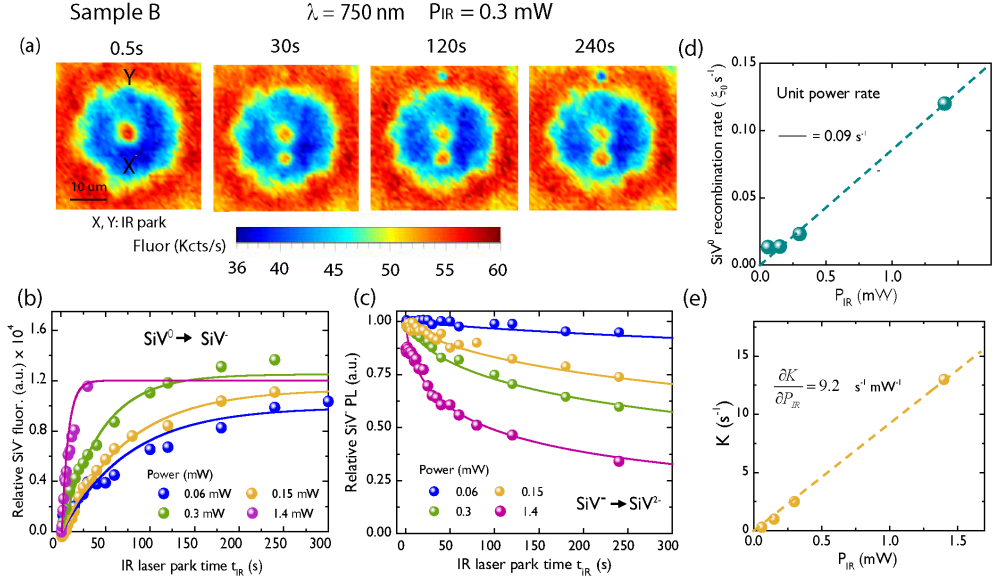

**Figure S9.** (a) Example of the confocal images upon application of the protocol in (a) for variable time in sample B. Each image is an averaged composite of 4 scans per park time to account for imperfect initialization and power drifts of the Ti:Sa laser beam (5%). (c) Integrated SiV$^-$ fluorescence (solid circles) at point X as a function of park time for variable laser powers under 750 nm excitation; solid lines are exponential fits. (c) Integrated SiV$^-$ fluorescence (solid circles) versus time at point Y; solid lines represent fits to the model of nitrogen-assisted electron tunneling (see section 3 of the supplementary material). In (b) and (c), a.u.: arbitrary units. (d) SiV$^0$ recombination rates and unit power rate extracted from the exponential fits show in (b). (c) coefficient $K$ versus power obtained from the fit of the SiV$^-$ to SiV$^{2-}$ recombination upon the model of nitrogen-assisted electron tunneling.

## 5. Computational methods

All density functional theory (DFT) calculations presented herein are performed within the PAW method[4,5], using Perdew-Burke-Ernzerhoff (PBE)[6] and hybrid (range-separated) Heyd-Scuseria-Ernzerhoff (HSE06)[7] functionals to account for electronic exchange-correlation interactions during atomic relaxations and self-consistent (SCF) calculations, respectively. The potential-energy surfaces (PES) of the various processes considered herein are all derived within the adiabatic Born-Oppenheimer approximation[8-10]. For all excited-state PES, the atoms are displaced along the configurational coordinate while keeping constrained occupation of the electronic states via the so-called constrained DFT (cDFT) method[9,10]. For the plane-wave basis, we use a low kinetic-energy cut-off of 370 eV as employed in previous work[11], which has shown to yield relatively well-converged results. All equilibrium defect structures are obtained by embedding the necessary impurities/vacancies in a 4×4×4 (512-atom) diamond supercell (created from a volume-optimized diamond unit-cell) and relaxing the ions at constant volume until forces are below $10^{-3}$ eV/Å. The electronic loops are converged down to energy differences below $10^{-8}$ eV (precision of wave functions) for equilibrium configurations, and $10^{-4}$ for non-equilibrium (PES) configurations. All supercell calculations are employed through sampling of the Brillouin zone at the $\Gamma$-point only. In obtaining the equilibrium charge-state transition energies, and all related calculations involving charged defects, we employ the charge correction scheme proposed in Ref. [12] to alleviate the obtained total energies from supercell finite-size effects. To calculate wave function overlaps between different defect centers, we make use of independent, large (2474-atom) diamond supercells for each center, with the electronic loops converged to $10^{-9}$ eV energy-differences so as to ensure well-converged 'tails' (low-amplitude regions) of the defect wave functions[11].

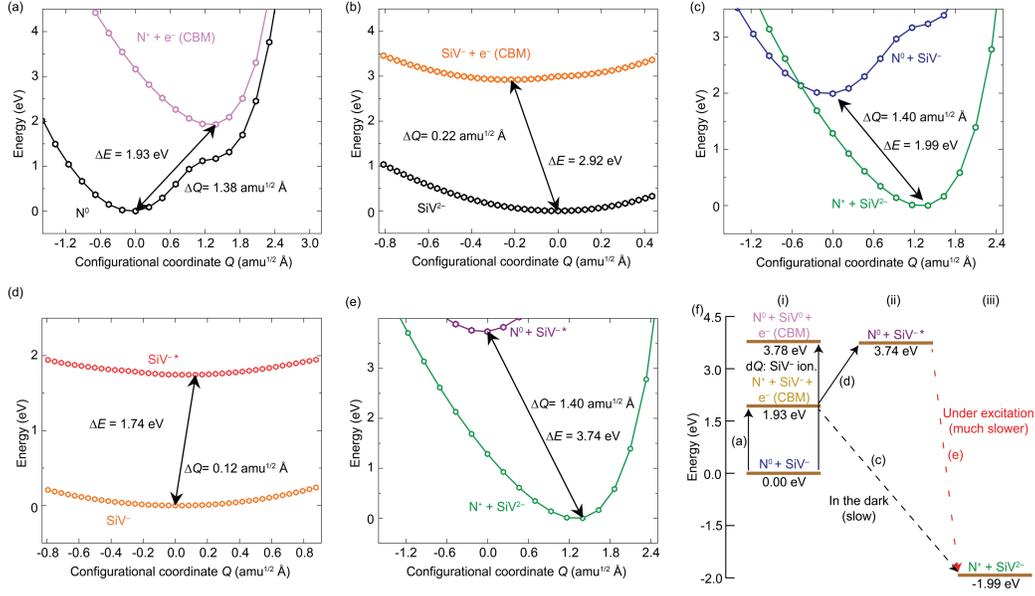

**Figure S10. Charge transfer involving the $N^0$ donor electron.** (a) Ionization of $N^0$ from the HOMO (via CBM). (b) Ionization of $SiV^{2-}$ via the CBM. (c) CCD for the combined individual processes depicted in (a) and (b). Here, the energetic contribution from the $e^-$ in the CB vanishes, putting the $N^+ + SiV^{2-}$ back at a lower energy compared to $N^0 + SiV^-$. (d) Optical excitation of $SiV^-$. (e) CCD for the combined individual processes depicted in (a) and (d). The energy difference between (c) and (e) correspond to the ZPL of $SiV^-$. (f) Energy-level diagram derived from the PES minima in (a–e), with the additional process of $SiV^-$ ionization into the CB, included as a reference (the corresponding CCD is not shown for brevity). All levels are given with respect to the $N^0 + SiV^-$ configuration, which is taken as the initial state for the charge transfer process.

## 6. Configurational coordinate diagrams and ionization/capture via the HOMO of $N^0$

This section considers alternative electron transfer pathways involving the HOMO states of $N^0$ and $SiV^-$. In Fig. S10, we show the full set of configurational coordinate diagrams (CCDs) with the transfer modelled by placing the electron in the diamond conduction band and adding the corresponding energy to the total energy of the system. In Fig. S10a, the calculated ionization threshold of $N^0$ lies well within the range of experimental values reported for such a process (1.7-2.2 eV)[13,14]. We can also see that such ionization from HOMO causes a substantial reconfiguration of the N-defect structure. In contrast, the reconfiguration for the ionization of $SiV^{2-}$ (or conversely, the electron capture by $SiV^-$ via the conduction band) remains much smaller (Fig. S10b). Thus, in a recombination process involving the $N^0$ and $SiV^-$ HOMOs (Fig. S10c), the rate will likely be dominated by the reconfiguration of $N^0$. For this process, we see that the PES curves cross, suggesting that electron transfer may happen in the dark (i.e., no $SiV^-$ excitation required). However, because the reconfiguration is rather large, this process is likely phonon-mediated and slow. Additional experimental work will be required to see whether formation of $SiV^{2-}$ via electron tunneling without light excitation can be seen experimentally; we note, however, that the localized nature of the donor electron orbital as compared to those of HOMO-1 and HOMO-2 would make this process more inefficient (see Fig. 4 and related paragraph in the main text).

In Fig. S10d, we show the CCD for the optical excitation of SiV¯, which (given its small experimental Huang-Rhys factor[15]), also has a small nuclear reconfiguration. Considering the SiV¯ in the excited state (i.e., (SiV¯)*) will shift the $N^0$ + SiV¯ curve in Fig. S10c by the Si⁻ ZPL energy (Fig. S10e), making the charge-transfer process even less likely. The resulting energy-level diagram summarizing the results in Fig. S10a-e is shown in Fig S10f. Here, the additional process of SiV¯ ionization is also included (as a reference for comparison). With these results, we conclude that while charge-transfer from $N^0$ to SiV¯ via the $N^0$ HOMO is plausible, it cannot explain the experimental observations discussed in the main text.